\documentclass[english,english,aps,pre,superscriptaddress,twocolumn,longtable]{revtex4}
\usepackage[T1]{fontenc}
\usepackage{graphicx}
\usepackage{amssymb}

\makeatletter

\providecommand{\tabularnewline}{\\}

\usepackage{graphicx}
\usepackage{ae}
\usepackage{epsfig}
\usepackage{graphicx}
\usepackage{dcolumn}
\usepackage{bm}
\usepackage{babel}
\makeatother
\begin{document}

\title{Orbital-Free Density Functional Theory: Kinetic Potentials and \textit{Ab
Initio} Local Pseudopotentials}

\author{Jeng-Da Chai%
\footnote{E-mail: jdchai@berkeley.edu. Present address: Molecular Foundry, 
Materials Sciences Division, Lawrence Berkeley National Laboratory, and
Department of Chemistry, University of California, Berkeley, California 94720 %
}}

\affiliation{Institute for Physical Science and Technology,}

\author{John D. Weeks}

\affiliation{Institute for Physical Science and Technology,}

\affiliation{and Department of Chemistry and Biochemistry, University of Maryland,
College Park, MD 20742}

\date{\today{}}

\begin{abstract}
In the density functional (DF) theory of Kohn and Sham, the kinetic
energy of the ground state of a system of noninteracting electrons
in a general external\label{abs:In-the-density}\label{abs:In-the-density}
field is calculated using a set of orbitals. Orbital free methods
attempt to calculate this directly from the electron density
by approximating the universal
but unknown kinetic energy density functional. However simple local
approximations are inaccurate and it has proved very difficult to
devise generally accurate nonlocal approximations. We focus instead
on the kinetic potential, the functional derivative of the kinetic
energy DF, which appears in the Euler equation for the electron density.
We argue that the kinetic potential is more local and more amenable to simple physically
motivated approximations in many relevant cases, and describe
two pathways by which the value of the kinetic energy can be efficiently calculated.
We propose two nonlocal
orbital free kinetic potentials that reduce to known exact forms
for both slowly varying and rapidly varying perturbations and also
reproduce exact results for the linear response of the density of
the homogeneous system to small perturbations. A simple and systematic
approach for generating accurate and weak \textit{ab initio} local
pseudopotentials which produce a smooth slowly varying valence component
of the electron density is proposed for use in orbital free DF calculations
of molecules and solids. The use of these local pseudopotentials further
minimizes the possible errors from the kinetic potentials.
Our theory yields results for the total energies and
ionization energies of atoms, and for the shell structure in the atomic
radial density profiles that are in very good agreement with calculations
using the full Kohn-Sham theory. 
\end{abstract}
\maketitle
\newpage

\section{Introduction}

Density-functional theory (DFT) has become one of the most powerful
tools for investigating the electronic structure of large complex systems.
In principle, as shown by Hohenberg and Kohn
\cite{Hohenberg}, the exact ground state energy of a system of $N$
electrons can be formally written as a functional $E[\rho]$ of only the
electron density $\rho({\bf r})$, a function of three variables,
and the external field $V_{ext}({\bf r})$.
Determining the energy and other ground state properties from such an approach
could dramatically reduce the computational cost for large systems when compared with traditional
quantum chemistry methods, which deal with wavefunctions
involving coordinates of all $N$ electrons~\cite{Parr,Dreizler}. 

Kohn and Sham (KS) \cite{Kohn,Sham} showed that $E[\rho]$ can be
usefully partitioned into the following set of terms:
\begin{equation}
E[\rho]=T_{s}[\rho]+E_{H}[\rho]+E_{xc}[\rho]+\int\rho({\bf r})V_{ext}({\bf r})d{\bf r}.
\label{eq:1}
\end{equation}
Here $T_{s}[\rho]$ is the \emph{noninteracting kinetic energy density
functional} (KEDF), which gives the kinetic energy of a model system
of $N$ noninteracting electrons in a self-consistent field chosen
so that the ground state density equals $\rho({\bf r})$,
\begin{equation}
E_{H}[\rho]\equiv\frac{1}{2}\int\int\frac{\rho({\bf r})\rho({\bf r'})}{|{\bf r}-{\bf r'}|}d{\bf r}d{\bf r'}
\label{eq:1a}
\end{equation}
is the classical \emph{electron-electron potential energy} (Hartree
energy) and $E_{xc}[\rho]$ is the \emph{exchange-correlation energy}
(including the difference between the interacting and noninteracting
kinetic energy and the difference between the quantum and classical
electron-electron potential energy). The last term on the right of
Eq.\ (\ref{eq:1}) is the only term that depends explicitly on the
external potential $V_{ext}({\bf r})$. Atomic units are used throughout
the paper.

If all these functionals were known, then the density $\rho({\bf r})$
could be obtained from the variational principle (Euler equation)
associated with minimizing Eq.\ (\ref{eq:1}): 
\begin{equation}
\mu=V_{T_{s}}({\bf r};[\rho])+V_{eff}({\bf r};[\rho]),\label{eq:2}
\end{equation}
and the total energy of the inhomogeneous system could then be determined
from the energy functional $E[\rho]$. All other physical quantities
related to the ground-state density could also be computed. Here $\mu$
is the chemical potential (the Lagrange multiplier associated with
the normalization condition $\int\rho({\bf r})d{\bf r}=N$), and $V_{eff}({\bf r};[\rho])$
is an effective one-body potential defined by 
\begin{eqnarray}
V_{eff}({\bf r};[\rho]) & \equiv & \frac{\delta}{\delta\rho({\bf r})}\left\{ E_{H}[\rho]+E_{xc}[\rho]
+\int\rho({\bf r})V_{ext}({\bf r})d{\bf r}\right\}\nonumber \\
 & = & V_{H}({\bf r};[\rho])+V_{xc}({\bf r};[\rho])+V_{ext}({\bf r}),\label{eq:4}
\end{eqnarray}
 where 
\begin{equation}
V_{H}({\bf r};[\rho])\equiv\delta E_{H}[\rho]/\delta\rho({\bf r})
=\int\frac{\rho({\bf r'})}{|{\bf r}-{\bf r'}|}d{\bf r'}\label{eq:4a}
\end{equation}
is the \emph{Hartree potential}, and $V_{xc}({\bf r};[\rho])\equiv\delta E_{xc}[\rho]/\delta\rho({\bf r})$
is the \emph{exchange-correlation potential}. Similarly we interpret

\begin{equation}
V_{T_{s}}({\bf r};[\rho])\equiv\delta T_{s}[\rho]/\delta\rho({\bf r})\label{eq:3}
\end{equation}
 as the \emph{kinetic potential} (KP)\ arising from the KEDF \cite{KingHandy}.

Further progress requires an accurate determination of
the noninteracting kinetic energy,
whose magnitude is much larger than the exchange-correlation energy.
The initial development of DFT as a practical computational method was made possible
by KS's realization that the numerical value of the noninteracting
kinetic energy can be exactly calculated, not directly from the density
itself using $T_{s}[\rho]$, but by introducing a set of $N$ one-electron
wave functions (orbitals) satisfying the $N$ coupled KS equations
that describe the model system \cite{Kohn,Sham}. 

Research could then focus
on determining the remaining small term $E_{xc}[\rho]$. Here even local density
approximations have often proved useful. Through
the efforts of many workers we now have generally accurate expressions
for $E_{xc}[\rho]$. Using these along with the KS orbitals to calculate the kinetic energy,
one can accurately calculate both the total energy $E[\rho]$ and the ground-state density
$\rho({\bf r})$ for a wide variety of systems.

However, the use of the KS orbitals usually generates
a relatively expensive $O(N^{3})$ scaling
of computational cost with the number of electrons. While this scaling is
much better than that of most standard methods that include
correlation energy, calculations for large systems remain problematic.
This remaining bottleneck could be removed if there were
an accurate treatment of the kinetic energy
in terms of the electron density only \cite{Parr,Dreizler,ChaiWeeks}.

To that end there has been considerable effort invested in developing
``orbital-free" density functional theory (OF-DFT) by making direct approximations for
$T_{s}[\rho]$ \cite{CAT,GAC,GAC2,WT,WGC,WGC2,Zhou2,Madden,Perrot,Alonso,Kaxiras,Herring,WC}.
While earlier simple local density approximations for $T_{s}[\rho]$ like those used in the
Thomas-Fermi (TF)\ model \cite{Thomas} are very inaccurate,
there have been two main advances in recent work that offer
prospects for significant improvements.

The first is the introduction of \emph{nonlocal}
KEDFs that reproduce known exact results for very slowly varying
or very rapidly varying fields and give the exact linear response
(LR) of the density of the uniform model system to small perturbations.
Similar ideas have been successfully applied to classical
nonuniform fluids \cite{Evans}.
The second advance is to focus not on the total density but on the
smaller and more slowly varying valence electron density as described
by a weak pseudopotential acting only on the valence electrons.
While conventional pseudopotential methods use orbitals,
recently developed \textit{ab initio local pseudopotential} (AILPS)
methods determine the unique local one-body potential producing
a given target valence density by solving the KS equations inversely,
using the one-to-one mapping between density and potentials in
DFT \cite{WangParr}. For OF-DFT with  LR-based KEDF's,
the use of pseudopotentials not only can reduce the computational
cost, but also can improve its accuracy, since the system will be closer to the LR
regime where $T_{s}[\rho]$ is designed to be accurate. \cite{Zhou,BWang}.
Indeed, very promising results using such OF-DFT methods
have been obtained for a variety of nearly free-electron-like metals.

However, existing KEDF's
have not yet achieved chemical accuracy for systems with localized
and more rapidly varying electron densities like molecules or for covalent or ionic
solids.  The main problem is that the exact
$T_{s}[\rho]$ is highly nonlocal, and we have little idea of the functional form
of the nonlocality for densities far from the LR regime.
It has proved very difficult to understand what errors an approximate
nonlocal $T_{s}[\rho]$ will produce in the density
as determined by the Euler equation with a general $V_{ext}({\bf r})$.

We explore here a different way to attack this basic problem.
The exact $T_{s}[\rho]$ can be formally obtained from
$V_{T_{s}}({\bf r};[\rho])$ by functional integration
over density changes in all regions of space \cite{Herring,KingHandy}.
Because of this integration $T_{s}[\rho]$ is a more nonlocal functional
of the density than is $V_{T_{s}}({\bf r};[\rho])$.
More detailed arguments arriving at this same conclusion have been
recently presented \cite{Holas}. Since most problems in
devising accurate approximations for $T_{s}[\rho]$ have arisen from
the nonlocality, this suggests it could be worthwhile to try to develop
approximations for the KP $V_{T_{s}}({\bf r};[\rho])$ itself.

To illustrate this point, Chai and Weeks \cite{ChaiWeeks} added a simple gradient
correction to the original local TF KP for atoms \cite{Thomas}, with
a coefficient chosen to reproduce the exact boundary condition of
exponential decay of the electron density far from the nucleus. 
Though quantitative results were not obtained, the
resulting modified Thomas-Fermi (MTF) model
gave energies for atoms and for closed-shell diatomic molecules that
showed notable improvements when compared to the original TF and related
gradient corrected KEDF models. However, the local gradient correction
used in the MTF KP cannot
reproduce the oscillatory atomic shell structure and
it does not satisfy the exact LR behavior in the
homogeneous limit. It is clear that nonlocality even in the KP must be taken
into account to achieve more accurate results.

We propose here new nonlocal approximations
for the KP using ideas similar to those employed for the nonlocal  KEDFs.
These new KPs satisfy the exact LR condition in the
uniform limit, and reproduce known exact limiting forms of
$V_{T_{s}}({\bf r};[\rho])$ both for very slowly varying and very rapidly
varying perturbations. As will become clear, the nonlocality in our
KP is determined by the requirement that LR
is exactly satisfied, and it is much easier to ensure that LR
holds for the KP than it is for analogous KEDF models.
We believe this level of nonlocality in the KP may suffice in many
cases when used in conjunction with AILPS methods to describe
slowly varying valence density components closer to the LR regime.

The remainder of this paper is organized as follow. Section II will
discuss some general pathways connecting
$T_{s}[\rho]$ and $V_{T_{s}}({\bf r};[\rho])$.
Section III will describe limiting forms of the KEDF and KP
for slowly varying and rapidly varying perturbations, and discuss LR theory,
an exact theory for the response of density of the uniform
electron gas to small perturbations. Section IV will develop
two nonlocal KPs incorporating both the
correct limiting forms of the exact KP and the exact LR of the free-electron
gas. Section V will compare the numerical results of the present method
for atoms with the KS-DFT and other KEDFs, both for all-electron calculations
and for valence electrons using the AILPS. We find that
the use of AILPS indeed reduces errors arising from nonlocality
in these approximate KPs or KEDFs, which give very accurate results
for the relatively slowly varying valence densities. Our conclusions
are given in Section VI.

\section{Pathways from $V_{T_{s}}({\bf r};[\rho])$ to $T_{s}[\rho]$}

If $T_{s}[\rho]$ is known, $V_{T_{s}}({\bf r};[\rho])$ can be simply
computed by functional differentiation. However, there is no unique
way of determining $T_{s}[\rho]$ from a given $V_{T_{s}}({\bf r};[\rho])$.
Many possible pathways can be used to construct $T_{s}[\rho]$ by
functional integration of $V_{T_{s}}({\bf r};[\rho])$  \cite{Pratt,Chen}.
If the exact $V_{T_{s}}({\bf r};[\rho])$ is used and
the integration is carried out exactly,
then all pathways would give the same exact result for $T_{s}[\rho]$.
However, when an approximate $V_{T_{s}}({\bf r};[\rho])$ is used,
different pathways will give different results for the kinetic energy.
But this ``thermodynamic inconsistency'' is small if reasonably
good approximations are used, since the integration
tends to smooth out local errors that may exist in the density \cite{ChaiWeeks,Chen}.

More problematic is the fact that most pathways require additional
results for partially coupled systems as the external field or density
perturbation is gradually turned on, which adds to the computational burden.
In particular, most earlier work has used a ``potential energy pathway'', where
the external potential is scaled by a coupling parameter \cite{ChaiWeeks,Pratt}.
The kinetic energy can then be found by subtracting the potential
energy (calculated from the potential energy density functionals)
from the total energy. However, this pathway is expensive,
since one has to solve the Euler
equation (\ref{eq:2}) for each partially coupled $V_{ext}^{\lambda}({\bf r})$
(with the same $\mu$), to determine the corresponding $\rho_{\lambda}$.

\subsection{Herring's Pathway}
However, Herring showed there is a particular pathway
arising from exact scaling relations between the non-interacting kinetic
energy $T_{s}[\rho]$ with respect to the coordinate ${\bf r}$ in
$\rho({\bf r})$ where very simple results involving only the final
density can sometimes be found \cite{Herring,KingHandy}.
If the coordinate ${\bf r}$ is scaled to $\alpha{\bf r}$,
the normalized scaled density is $\rho_{\alpha}({\bf r})=\alpha^{3}\rho(\alpha{\bf r})$.
It is easy to show that the exact $T_{s}[\rho]$ then obeys 
\begin{equation}
\alpha^{2}T_{s}[\rho]=T_{s}[\rho_{\alpha}].\label{eq:5}
\end{equation}
 For isolated systems, such as atoms and molecules, the density and
its derivatives to all order vanish far from the nuclei. For such systems,
when Eq.\ (\ref{eq:5}) is differentiated with respect to $\alpha$,
and the partial derivative is evaluated at $\alpha=1$, we find the
formally exact result 
\begin{equation}
T_{s}[\rho]=\frac{1}{2}\int V_{T_{s}}({\bf r};[\rho])\nabla\cdot({\bf r}\rho({\bf r}))d{\bf r}.\label{eq:6}
\end{equation}

Therefore, once the kinetic potential $V_{T_{s}}({\bf r};[\rho])$
is known for some given $\rho({\bf r})$, the numerical value of $T_{s}[\rho]$
can then be immediately determined from Eq.\ (\ref{eq:6}). Since there is no need to perform
a coupling parameter integration over the change of density or potential,
this scheme is not only fast, but also numerically reliable. The final
form of Eq.\ (\ref{eq:6}) is essentially the viral theorem,
and is directly related to the force on molecules \cite{PerdewKurth,Sham2,Levy,Averill}. 

Note that this simple and exact pathway holds only for the \emph{noninteracting}
$T_{s}[\rho]$ \cite{Levy}, which again shows the virtues of the
KS partitioning of the total energy. We will use Eq.\ (\ref{eq:6})
as the basic pathway to determine the numerical value of $T_{s}[\rho]$
from a given approximate $V_{T_{s}}({\bf r};[\rho])$ for most calculations
in this paper. However, Eq.\ (\ref{eq:6}) does not hold for extended solid-state
systems because of nonvanishing boundary terms, and thus far
we have not found an exact and simple way of including them.

Fortunately, there is another class of computationally efficient
``density pathways'' that can be used for extended systems,
as we now show. Density pathways can also be used for atomic and molecular systems
to check the accuracy of the $V_{T_{s}}$ used,
since results using the exact $V_{T_{s}}$ would
be independent of path \cite{Pratt,Chen}.

\subsection{Density Pathways}

The change in the kinetic energy can be formally related to a coupling
parameter integration, where the density changes from some known value
at $\lambda=0$ to the final density at $\lambda=1$:
\begin{equation}
T_{s}[\rho]=T_{\lambda=0}+\int_{0}^{1}d\lambda\int d{\bf r}V_{T_{s}}({\bf r};[\rho_{\lambda}])
\frac{\partial\rho_{\lambda}({\bf r})}{\partial\lambda}\label{eq:2.3.1}
\end{equation}
In most cases a simple linear density pathway will suffice.  Here
the density $\rho({\bf r})$ is linearly scaled by a coupling parameter $\lambda$
from some uniform reference density $\rho_{0}$ naturally chosen
to be the uniform electron density $N/V$ in extended systems:
\begin{equation}
\rho_{\lambda}({\bf r})=\rho_{0}+\lambda[\rho({\bf r})-\rho_{0}].
\label{eq:2.3.2}
\end{equation}
Then Eq.\ (\ref{eq:2.3.1}) becomes
\begin{equation}
T_{s}[\rho]=T_{\lambda=0}+\int_{0}^{1}d\lambda\int d{\bf r}V_{T_{s}}({\bf r};[\rho_{\lambda}])[\rho({\bf r})-\rho_{0}]
\label{eq:2.3.4}
\end{equation}

Here $T_{\lambda=0}$ is the kinetic
energy of the uniform system, i.e., the Thomas-Fermi kinetic
energy $T_{TF}[\rho_0]$. For extended systems, where the Herring's pathway
cannot be used, this density pathway appears to be a good way to compute $T$.
Other density pathways, like the square-root pathway introduced by Chen and
Weeks \cite{Chen} to describe nonuniform hard sphere fluids,
can be defined, and have proved useful in certain
applications, but we do not consider them here.

Note from Eq.\ (\ref{eq:2.3.2}) that $\rho_{\lambda}({\bf r})$
depends only on the final
density, so evaluation of Eq.\ (\ref{eq:2.3.4}) is straightforward
and this pathway is computationally efficient.
Unlike the potential energy pathway where the external
potential is scaled,
there is no need to solve the Euler equation (\ref{eq:2}) for
its corresponding external potential $V_{ext}^{\lambda}({\bf r})$
at each $\lambda$.
However, for isolated systems, where
$\rho_{0}=0$, this pathway is likely to be less accurate than
Herring's pathway, since it does not automatically satisfy the virial theorem. 

\section{Exact Limits and Linear Response Theory}

Although the exact $T_{s}[\rho]$ is still unknown, several limiting
forms have been discovered for particular density distributions.
These provide important cornerstones that can be used
to construct
accurate KEDFs and KPs in many cases, as will be seen below.

In particular, the Thomas-Fermi (TF) KEDF
\cite{Thomas} is known to be exact for a uniform system: 
\begin{equation}
T_{TF}[\rho]=C_{F}\int\rho^{5/3}({\bf r})d{\bf r},\label{eq:7}
\end{equation}
 where $C_{F}=\frac{3}{10}(3\pi^{2})^{2/3}$. The TF KEDF $T_{TF}[\rho]$
is derived by local use of uniform free-electron gas model, and is
exact for a system with an infinite number of electrons. The corresponding expression
for the TF KP is
\begin{equation}
V_{TF}({\bf r};[\rho])\equiv\delta T_{TF}[\rho]/\delta\rho({\bf r})=\frac{5}{3}C_{F}\rho^{2/3}({\bf r}).
\label{eq:14a}
\end{equation}
This depends only on the local value of $\rho^{2/3}({\bf r})$ and thus formally is
more local than the TF KEDF, whose functional dependence on $\rho$
involves the density at all $\bf r$. Of course in this
simple case the functional integration of Eq.\ (\ref{eq:14a}) can be carried out exactly to yield
Eq.\ (\ref{eq:7}), but this cannot be done in general and the nonlocality of $T_s[\rho]$ has proved problematic.

Results for nonuniform systems are best described in Fourier space.
For a very slowly varying perturbation of the density, the second-order
gradient expansion is exact \cite{Yang}. It is easy to see that results correct to second order
at small wavevectors are given by
\begin{equation}
T_{TF\frac{1}{9}W}[\rho]\equiv T_{TF}[\rho]+\frac{1}{9}T_{W}[\rho],\label{eq:8}
\end{equation}
 where 
\begin{equation}
T_{W}[\rho] \equiv \frac{1}{8}\int\frac{|{\bf \bigtriangledown}\rho({\bf r})|^{2}}{\rho({\bf r})}d{\bf r}\label{eq:8a}
\end{equation}
is the von Weizs\"{a}cker (W) KEDF \cite{Weizsacker}.

$T_{W}[\rho]$ is exact for a system with one or two electrons,
or where the density can be accurately described by a single orbital.
Moreover it has been argued \cite{CAT,WGC,WGC2,WC}
that $T_{W}[\rho]$ gives the correct leading order term for a rapidly varying
perturbation with only high wavevector components and that the next order
correction is reproduced by 
\begin{equation}
T_{W-\frac{3}{5}TF}[\rho]\equiv T_{W}[\rho]-\frac{3}{5}T_{TF}[\rho].\label{eq:9}
\end{equation}

The W KP is
\begin{equation}
V_{W}({\bf r};[\rho])\equiv\delta T_{W}[\rho]/\delta\rho({\bf r})=
\frac{1}{8}\left(\frac{|{\bf \bigtriangledown}\rho({\bf r})|^{2}}{\rho^{2}({\bf r})}
-2\frac{{\bf \bigtriangledown}^{2}\rho({\bf r})}{\rho({\bf r})}\right).\label{eq:14b}
\end{equation}
If we represent the full density by an effective single
orbital function $\psi({\bf r})$, 
\begin{equation}
\rho({\bf r})=|\psi({\bf r})|^{2}\label{eq:29aa}
\end{equation}
then the W KP can be written in a compact form that will later prove useful:
\begin{equation}
V_{W}({\bf r};[\rho])=-\frac{{\bf \bigtriangledown}^{2}\psi({\bf r})}{2\psi({\bf r})}.\label{eq:29ab}
\end{equation}

Finally, the linear response of the density of a uniform non-interacting
electron gas with density $\rho_{0}$ to a small perturbation
$\delta V({\bf k})=\epsilon_{{\bf k}}e^{i{\bf k}\cdot{\bf r}}$
is exactly known \cite{Lindhard}, 
\begin{equation}
\delta\rho({\bf k})=\chi_{L}(q)\delta V({\bf k}).\label{eq:11}
\end{equation}
Here
\begin{equation}
q\equiv k/2k_{F} \label{eq:11a}
\end{equation}
is a dimensionless wavevector,
where
\begin{equation}
k_{F}\equiv(3\pi^{2}\rho_{0})^{1/3}\label{eq:11b}
\end{equation}
is the \emph{Fermi wavevector} (FWV) and $k\equiv|{\bf k}|$.
The LR function $\chi_{L}(q)$ has the form
\begin{eqnarray}
\chi_{L}(q) & = & -\frac{k_{F}}{\pi^{2}}F_{L}^{-1}(q)\nonumber \\
 & = & -\frac{k_{F}}{\pi^{2}}\left[\frac{1}{2}+\frac{1-q^{2}}{4q}\ln\left|\frac{1+q}{1-q}\right|\right],\label{eq:12}
\end{eqnarray}
where 
\begin{equation}
F_{L}(q)\equiv\left[\frac{1}{2}+\frac{1-q^{2}}{4q}\ln\left|\frac{1+q}{1-q}\right|\right]^{-1}\label{eq:12a}
\end{equation}
has been called the Lindhard function \cite{WC}.

It is known that the weak logarithmic singularity at $q=1$ in $F_{L}^{-1}(q)$
is responsible for Friedel oscillations, and may also be important
for the appearance of atomic shell structure. This singularity further
divides the Lindhard function into two branches in Fourier space:
the low-momentum ($q<1$ ) or the low-$q$ (LQ) branch, and the high-momentum
($q>1$) or the high-$q$ (HQ) branch \cite{WC}.

The dimensionless
response function arising from the TF KEDF is $F_{TF}(q)=1$, and that from
the W KEDF is $F_{W}(q)=3q^{2}$ \cite{Jones}. Clearly, no linear
combination of the TF and the W KEDFs can reproduce the exact Lindhard
function in Eq.\ (\ref{eq:12a}). This has the following two limits
\cite{WC}, 
\begin{equation}
F_{L}(q)=\left\{ \begin{array}{ll}
1+q^{2}/{3}+O(q^{4}) & q\ll1\\
3q^{2}-3/{5}+O(q^{-2}) & q\gg1\end{array}\right.\label{eq:13}
\end{equation}

It should be noted that the expansions for both the low-$q$ and high-$q$
limits are correct to all orders in perturbation theory, but valid
only in the appropriate limits in Fourier space. On the other hand,
the LR theory is valid for all wavevectors, but is only accurate for
small perturbations. Therefore, the regime where the
response functions of the two limiting KEDFs deviate from the exact
LR function gives an indication of the range of wavevectors where the two limiting
forms are inaccurate.

\begin{figure}

\caption{\label{Fig.1}Linear response functions of a uniform system of noninteracting
Fermions as given by the TF, W, and MTF (see Ref.\ \cite{ChaiWeeks}) models.}

\includegraphics[%
  scale=0.7]{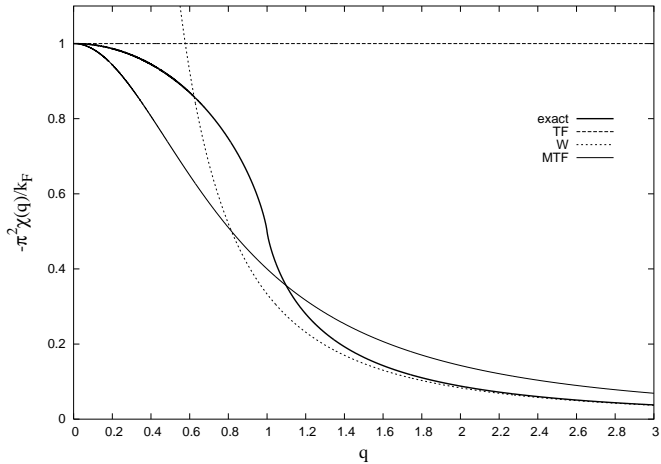}
\end{figure}

As shown in Fig.\ \ref{Fig.1}, the response function 
$F_{TF}^{-1}(q)$ has no momentum dependence and is only exact at $q=0$.
The response function $F_{W}^{-1}(q)$ is exact asymptotically
at high $q$, and remains fairly accurate for $q\gtrsim2$, but
is divergent in the low-$q$ branch, and fails completely for the
nearly uniform electron gas. In contrast, the MTF model \cite{ChaiWeeks}
gives a reasonably accurate average description of the exact response function,
especially in the important region near the singularity at $q=1$.

\section{Construction of Nonlocal Kinetic Potentials}
\subsection{Kinetic Energy Density Functionals $T_{TF\lambda W}[\rho]$ and $T_{W\lambda TF}[\rho]$}

Simple linear combinations of the two limiting KEDF's in Eqs.\ (\ref{eq:7}) and (\ref{eq:8a}),
such as the TF$\lambda$W KEDF \cite{Yang,Tomishima,Lieb,Chan}
\begin{equation}
T_{TF\lambda W}[\rho]\equiv T_{TF}[\rho]+\lambda T_{w}[\rho]\label{eq:13a}
\end{equation}
 and the W$\lambda$TF KEDF \cite{Acharya,Gazquez,Bartolotti,Acharya2}
\begin{equation}
T_{W\lambda TF}[\rho]\equiv T_{W}[\rho]+\lambda T_{TF}[\rho]\label{eq:13b}
\end{equation}
 have been widely studied for several decades. The value of the parameter
$\lambda$ was either determined empirically for getting good atomic
energy or obtained by some semiclassical arguments.

The advantage of these approaches is the ability to generate a family of simple
KEDF's easily. It has been shown empirically that the
$TF1/5W$ model can give good values for atomic energies, but
the predicted density profiles are generally
not very accurate, both near and far away from the nucleus.
The $T_{TF\lambda W}[\rho]$ and $T_{W\lambda TF}[\rho]$
functionals give the correct leading term in the density response to a slowly-varying
perturbation and a rapidly-varying perturbation respectively, and with particular
choices of $\lambda$ as in Eqs.\ (\ref{eq:8}) and (\ref{eq:9}) they can reproduce
the next order term. Unfortunately,
they then will have an incorrect leading term in the opposite limit, unless
$\lambda=1$. However it has been shown that $T_{TFW}[\rho]$ with
$\lambda=1$ always overestimates the exact $T_{s}[\rho]$ for various
systems \cite{WC}. Finally, none of these functionals can
reproduce the exact response function $F_{L}(q)$ in the homogeneous
limit. Since these models fail to satisfy all the known limits, and nonlocality
in $T_{s}[\rho]$ is not correctly described, it is also not surprising
that atomic shell structure is missing in these approaches.

\subsection{Combining TF and W Kinetic Potentials}

We argue that it may be more profitable to take advantage of known
limiting forms of the KP, rather than the KEDF, and
develop approximations for the more local $V_{T_{s}}({\bf r};[\rho])$
directly. Again we can rely
on known results in the linear response regime when the density variations
are not too large.

From Eqs.\ (\ref{eq:8}) and (\ref{eq:9}), the following linear combinations of
the TF KP and the W KP in Eqs.\ (\ref{eq:14a}) and (\ref{eq:14b})
can reproduce exact results to second order for very small and very
large wavevector perturbations respectively: 
\begin{equation}
V_{T_{s}}({\bf r};[\rho])\approx\left\{ \begin{array}{ll}
V_{TF}({\bf r};[\rho])+\frac{1}{9}V_{W}({\bf r};[\rho]) & \,\,\,\, q\ll1\\
V_{W}({\bf r};[\rho])-\frac{3}{5}V_{TF}({\bf r};[\rho]) & \,\,\,\, q\gg1\end{array}\right.\label{eq:22}
\end{equation}
 Since $V_{TF}({\bf r};[\rho])$ and $V_{W}({\bf r};[\rho])$ are
the only components up to second order of the two exact limiting forms
of the KP, we can combine them in analogy to the $TF\lambda W$ and
$W\lambda TF$ models and arrive at generalized KPs.

However, instead of combining them using a fixed parameter
$\lambda$, it seems natural to represent them in Fourier space
and allow a wavevector dependence in $\lambda=\lambda(q)$
to connect the limiting forms. The $\lambda(q)$ can then be chosen
in a very simple way so that the exact LR function is reproduced
for a uniform system with density $\rho_{0}$. In this way the LR
function \emph{bridges} the exact limits at large and small wavevectors,
and if the theory is applied to weak perturbations in the linear response
regime for intermediate wavevectors we can expect very accurate results.
Here, we derive such generalized KPs based on the KP for the $W\lambda TF$ model.

\subsection{HQ Kinetic Potential}

In analogy to the W$\lambda$TF
model in Eq.\ (\ref{eq:13b}) we look for a kinetic potential
of the form
\begin{eqnarray}
V_{HQ}^{0}({\bf k}) & = & V_{W}({\bf k})+\lambda_{HQ}(q)\,\, V_{TF}({\bf k})\nonumber \\
 & = & V_{TF}({\bf k})+V_{W}({\bf k})+\hat{f}(q)\
 V_{TF}({\bf k}),\label{eq:4.2.6}
 \end{eqnarray}
where
\begin{equation}
q=k/2k_{F}\label{eq:4.2.2}
\end{equation}
is a dimensionless wavevector normalized by the FWV $k_{F}$ in
Eq.\ (\ref{eq:11b}) of a uniform reference system with density $\rho_{0}$
and $\hat{f}(q)= \lambda_{HQ}(q)-1$.
The superscript $0$ in $V_{HQ}^{0}$ indicates use of a uniform
reference system. For a small perturbation, we can linearize the $V_{HQ}^{0}({\bf k})$
in Eq.\ (\ref{eq:4.2.6}). Requiring that it satisfy LR exactly then
determines the weight function $\hat{f}(q)$ as
 \begin{equation}
\hat{f}(q)=F_{L}(q)-3q^{2}-1.\label{eq:4.2.7}
\end{equation}
\begin{figure}

\caption{\label{Fig.4.1} Weight function $\hat{f}(q)$ for the HQ and LQ KPs.}

\includegraphics[%
  scale=0.7]{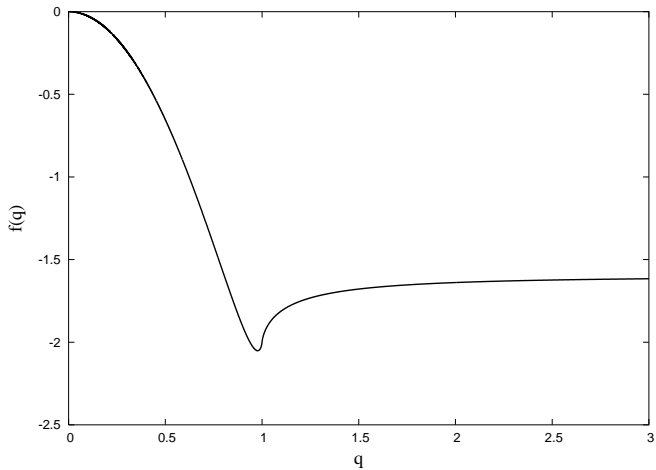}
\end{figure}
See Fig.\  \ref{Fig.4.1}.

We refer to Eq.\ (\ref{eq:4.2.6}) with Eq.\ (\ref{eq:4.2.7}) as the HQ KP model.
It reproduces the correct high-$q$ limit in Eq.\ (\ref{eq:22})
up to the second order.  However, unlike Eq.\  (\ref{eq:9}), it also
satisfies the correct low-$q$ limit to leading
order and gives exact results for all $q$ in the linear response regime.
Inverse Fourier transform of Eq.\ (\ref{eq:4.2.6}) then gives
\begin{equation}
\begin{array}{ll}
V_{HQ}^{0}({\bf r};[\rho],k_{F})=V_{TF}({\bf r};[\rho])+V_{W}({\bf r};[\rho])\textbf{\medskip}\\
\;\;\;\;\;\;\;\;\;\;\;\;\;\;+{\displaystyle \int f(|{\bf r}-{\bf r'}|;k_{F})\,\, V_{TF}({\bf r'};[\rho])d{\bf r'}.}
\end{array}\label{eq:4.2.8}
\end{equation}
 
This expression is directly useful for extended systems where a reasonable
$\rho_{0}$ can be defined.
For isolated systems such as atoms and molecules
where the density vanishes far from the nuclei, it seems natural to replace
$k_{F}$ in Eq.\ (\ref{eq:4.2.8}) by the \emph{local} Fermi wavevector (LFWV)
\begin{equation}
k_{F}({\bf r})\equiv(3\pi^{2}\rho({\bf r}))^{1/3},\label{eq:3.6.2}
\end{equation}
though errors may be introduced for rapidly varying density distributions.
Using Eq.\ (\ref{eq:14a}), this yields the general form of our proposed HQ kinetic potential:
\begin{equation}
\begin{array}{ll}
V_{HQ}({\bf r};[\rho],k_{F}({\bf r}))=V_{TF}({\bf r};[\rho])+V_{W}({\bf r};[\rho])\textbf{\medskip}\\
\;\;\;\;\;\;\;\;\;\;\;\;\;\;+\frac{5}{3}C_{F}{\displaystyle \int f(|{\bf r}-{\bf r'}|;k_{F}({\bf r}))\,\,\rho^{2/3}({\bf r'})d{\bf r'}.}
\end{array}\label{eq:4.2.9}
\end{equation}
Note that the last term in Eq.\ (\ref{eq:4.2.9}) is most easily computed
in Fourier space as 
\begin{equation}
\frac{5C_{F}}{3(2\pi)^{3}}\int\hat{f}(k/2k_{F}({\bf r}))\rho^{2/3}({\bf k})e^{-i{\bf k}\cdot{\bf r}}d{\bf k},
\label{eq:4.2.10}
\end{equation}

As can been seen in Eq.\ (\ref{eq:4.2.7}), the weight function $\hat{f}(k/2k_{F}({\bf r}))$
is determined analytically. Unlike the LR-based  KEDF approaches, no first-order
differential equation is needed to solve for the weight function in Fourier space. For atomic systems
where the LFWV is used, the convolution in Eq.\ (\ref{eq:4.2.10}) must be carried out numerically,
which will lead to a quadratic scaling of the HQ model (and related LQ model described below)
in the number $N$ of electrons.
For extended systems where one can expand about the local density $\rho_0$,
one can use fast Fourier transforms (FFT's) for a much more efficient 
computation of this integral \cite{Singh}. 

\subsection{LQ Kinetic Potential}

In analogy to the TF$\lambda$W model in Eq.\ (\ref{eq:13a}), we could similarly
generate a KP that is accurate to second order at low $q$ while still reproducing
the leading term at high $q$.
However this is numerically less useful because the analogue
of Eq.\ (\ref{eq:4.2.10}) involves the Fourier transform of 
$V_{W}({\bf r'};[\rho])$, which cannot be simply expressed in terms of the density.
Instead, by empirically taking a properly chosen component of the density
outside the integral we find that Eq.\ (\ref{eq:4.2.9}) can be
modified to produce a new LQ KP that
is accurate to second order at low $q$ and first order at high $q$ \cite{Chaidiss}:
\begin{equation}
\begin{array}{ll}
V_{LQ}({\bf r};[\rho],k_{F}({\bf r}))=V_{TF}({\bf r};[\rho])+V_{W}({\bf r};[\rho])\textbf{\medskip}\\
\;\;\;\;\;\;\;\;\;\;\;\;+\frac{20}{9}C_{F}\rho^{1/6}({\bf r}){\displaystyle \int f(|{\bf r}-{\bf r'}|;k_{F}({\bf r}))\,\,\rho^{1/2}({\bf r'})d{\bf r'}}
\end{array}\label{eq:4.3.5}
\end{equation}

Extending these ideas we have constructed a modified KP
that satisfies LR everywhere and is accurate to second order at both low and high $q$ 
\cite{Chaidiss}.
However the functional form is much more complicated, and
little additional accuracy is gained from the improved
behavior at very small or very large wavevectors,
since all forms use LR to interpolate for intermediate wavevectors, and this
is where most errors arise in practice.
Thus we will report results here only for the HQ and LQ models.

\section{Results for Atoms using the HQ and LQ Models}

For completeness and to compare to earlier work, we first briefly discuss
all-electron calculations using the proposed
HQ and LQ model KPs and the full atomic potentials. We then describe
results using AILPS methods. These are compared with the KS-DFT, the TF$\lambda$W
models, and the CAT model introduced by
Chac\'on, Alvarellos, and Tarazona \cite{CAT,GAC,GAC2}.
The CAT model is a LR-based KEDF method, which gives some indication of
shell structure. We employed the latest version, which uses a nonlocal
two-body Fermi wave vector with a prescribed functional form depending
on an empirical parameter $\beta=1/2$ (defined in Eq.\ (3) of
Ref.\ \cite{GAC2}). This caused the numerical calculations \cite{CATcalculations}
to be considerably more costly than
those of the LQ or HQ models, which used the local Fermi wave vector as in Eq.\ (\ref{eq:3.6.2}).
All calculations are spin-restricted and use the local density approximation
(LDA) \cite{Dirac,Ceperley,Perdew} for the exchange-correlation functional.

\subsection{All-electron calculations}

All-electron calculations consider the density response to the large and
rapidly varying nuclear potential. Since the system is far from the linear response regime,
quantitative results from
the HQ and LQ models (or from LR-based KEDF methods) cannot be expected.
However, by incorporating exact results for very large and very small wavevectors,
these models do correct major deficiencies of the purely local TF model
(which, e.g., predicts an infinite density at the nucleus!)
and even give some qualitative indications of atomic shell structure.

\begin{table}

\caption{\label{table:1}Atomic energy $E$ using 
the KS, LQ, HQ, CAT and the TF$\lambda$W models in all-electron calculations.
MAPE, the mean absolute percentage error (relative to the KS method) of various
OF models are given at the bottom of their respective columns.}

\begin{tabular}{ccccccc}
\hline 
&
 KS&
 LQ&
 HQ&
 CAT&
 TF1/5W&	
 TFW\tabularnewline
\hline
He&
 $-2.834$&
 $-2.565$&
 $-2.437$&
 $-2.675$&
 $-2.911$&
 $-1.559$\tabularnewline
\hline
Ne&
 $-128.2$&
 $-134.3$&
 $-126.6$&
 $-126.2$&
 $-129.5$&
 $-86.40$\tabularnewline
\hline
Ar&
 $-525.9$&
 $-545.9$&
 $-512.2$&
 $-515.1$&
 $-526.2$&
 $-375.5$\tabularnewline
\hline
Kr&
 $-2750$&
 $-2805$&
 $-2621$&
 $-2712$&
 $-2748$&
 $-2099$\tabularnewline
\hline
Xe&
 $-7229$&
 $-7306$&
 $-6844$&
 $-7141$&
 $-7214$&
 $-5701$\tabularnewline
\hline
Be&
 $-14.45$&
 $-14.39$&
 $-13.64$&
 $-14.11$&
 $-14.71$&
 $-8.699$\tabularnewline
\hline
Mg&
 $-199.1$&
 $-207.9$&
 $-195.7$&
 $-195.2$&
 $-200.0$&
 $-136.4$\tabularnewline
\hline
C&
 $-37.42$&
 $-38.97$&
 $-36.85$&
 $-37.25$&
 $-38.41$&
 $-24.01$\tabularnewline
\hline
N&
 $-54.02$&
 $-56.71$&
 $-53.59$&
 $-53.84$&
 $-55.39$&
 $-35.33$\tabularnewline
\hline
O&
 $-74.47$&
 $-78.39$&
 $-74.02$&
 $-74.08$&
 $-76.11$&
 $-49.39$\tabularnewline
\hline
Si&
 $-288.2$&
 $-300.4$&
 $-282.5$&
 $-282.2$&
 $-288.9$&
 $-200.5$\tabularnewline
\hline
P&
 $-339.9$&
 $-354.1$&
 $-332.7$&
 $-332.8$&
 $-340.6$&
 $-238.3$\tabularnewline
\hline
S&
 $-396.7$&
 $-412.8$&
 $-387.7$&
 $-388.5$&
 $-397.3$&
 $-279.9$ \tabularnewline
\hline
MAPE&
 &
 $4.06 \%$&
 $3.42 \%$&
 $1.82 \%$& 
 $1.10 \%$&
 $32.0 \%$ \tabularnewline
\hline

\end{tabular}
\end{table}

The numerical method use the \emph{Pauli kinetic potential} $V^{P}({\bf r};[\rho])$
\cite{March,Holas2}, defined as 
\begin{equation}
V^{P}({\bf r};[\rho])\equiv V_{T_{s}}({\bf r};[\rho]) - V_{W}({\bf r};[\rho]).
\label{eq:49a}
\end{equation}
Since $V_{W}({\bf r};[\rho])$ is the exact KP for a system where
the density can be accurately described by a single orbital, if $V^{P}({\bf r};[\rho])$
is omitted, one would essentially obtain the ground state density
of the corresponding Boson system, where all the electrons are in
the same orbital. If we represent the full density by a single orbital
function $\psi({\bf r})$, so that $\rho({\bf r})$ and $V_{W}({\bf r};[\rho])$
can be written in the forms of Eq.\ (\ref{eq:29aa}) and Eq.\ (\ref{eq:29ab})
respectively, we can then combine $V^{P}({\bf r};[\rho])$ with the
one-body potential $V_{eff}({\bf r};[\rho])$ in Eq.\ (\ref{eq:2}),
and derive a Schr\"odinger-like equation for the Bose orbital
$\psi({\bf r})$, 
\begin{equation}
\left\{ -{\textstyle \frac{1}{2}}{\bf \bigtriangledown}^{2}+V_{eff}({\bf r};[\rho])
+V^{P}({\bf r};[\rho])\right\} \psi({\bf r})=\mu\psi({\bf r}).
\label{eq:51}
\end{equation}
In other words, $V_{eff}({\bf r};[\rho])+V^{P}({\bf r};[\rho])$
is now the one-body effective potential for the corresponding Boson
system with the same electron density. This reduction of an $N$-fermion
problem to a Boson form is widely implemented in OF-DFT due to its
numerical stability and its easy implementation using existing KS-DFT
codes \cite{WC}.

The associated Pauli potentials for the HQ and LQ models
are immediately obtained by subtraction of the W KP from Eq.\ (\ref{eq:4.2.9})
and Eq.\ (\ref{eq:4.3.5}) respectively.
The standard finite difference method for solving Euler equations
for the TFW models \cite{Abrahams} are implemented for the LQ and
HQ models, and the nonlocal terms are evaluated by Fourier transforms.
The choices of radial grids for both of the real and Fourier space
and other detailed numerical methods are given in Ref.\  \cite{ChaiWeeks}.
The kinetic energy for HQ and LQ models is computed using the Herring
pathway in Eq.\  (\ref{eq:6}).

\begin{table}

\caption{\label{table:2}Electron density at the nucleus $\rho(0)$, using
the KS, LQ, HQ, CAT and the TF$\lambda$W models in all-electron calculations.
MAPE, the mean absolute percentage error (relative to the KS method) of various
OF models are given at the bottom of their respective columns.}

\begin{tabular}{ccccccc}
\hline 
&
 KS&
 LQ&
 HQ&
 CAT&
 TF1/5W&
 TFW\tabularnewline
\hline
He&
 $3.525$&
 $3.088$&
 $2.742$&
 $3.600$&
 $18.23$&
 $0.9515$\tabularnewline
\hline
Ne&
 $614.5$&
 $576.6$&
 $517.6$&
 $613.2$&
 $2596$&
 $169.6$\tabularnewline
\hline
Ar&
 $3819$&
 $3642$&
 $3282$&
 $3812$&
 $1.548\times10^{4}$&
 $1093$\tabularnewline
\hline
Be&
 $34.86$&
 $30.49$&
 $27.17$&
 $33.75$& 
 $158.2$&
 $8.952$\tabularnewline
\hline
Mg&
 $1086$&
 $1024$&
 $920.9$&
 $1083$&
 $4519$&
 $303.0$\tabularnewline
\hline
C&
 $126.0$&
 $113.3$&
 $101.2$&
 $122.8$& 
 $547.9$&
 $33.07$\tabularnewline
\hline
N&
 $203.9$&
 $185.6$&
 $166.1$&
 $200.0$&
 $876.7$&
 $54.24$\tabularnewline
\hline
O&
 $308.6$&
 $284.1$&
 $254.6$&
 $304.6$& 
 $1317$&
 $83.19$\tabularnewline
\hline
Si&
 $1754$&
 $1662$&
 $1495$&
 $1749$&
 $7218$&
 $493.9$\tabularnewline
\hline
P&
 $2173$&
 $2062$&
 $1857$&
 $2167$& 
 $8901$& 
 $614.5$\tabularnewline
\hline
S&
 $2654$&
 $2523$&
 $2272$&
 $2647$&
 $1.083\times10^{4}$&
 $753.5$ \tabularnewline
\hline
MAPE&
 &
 $7.61 \%$&
 $17.2 \%$&
 $1.14 \%$&
 $331 \%$&
 $72.6 \%$\tabularnewline
\hline

\end{tabular}
\end{table}

 As shown in Table \ref{table:1}, the atomic energy calculated by the energy-optimized
TF1/5W model is very close to the KS-DFT, and outperforms
all the LR-based models, and other TF$\lambda$W models.
In Table \ref{table:2}, we compare the electron density at the nucleus $\rho(0)$ for various
models. The TF1/5W model overestimates $\rho(0)$ by about a factor of 4,
while the TFW model underestimates it by about 30\%. The predicted
values of $\rho(0)$ for all the LR-based models are very close to the KS results, and are
much better than the TF$\lambda$W models. 

\begin{figure}[t]

\caption{\label{Fig.4} (Color online) Radial density $r^{2}\rho(r)$ of the Kr atom using the
KS method, the LQ and HQ models, the CAT model,
and the TF$\lambda$W models (see the inset) with
the full nuclear potential.}

\includegraphics[%
  scale=0.7]{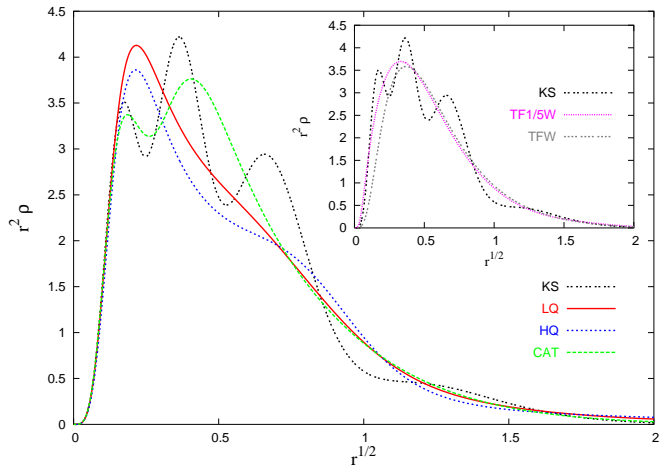}
\end{figure}

In Fig.\ (\ref{Fig.4}), we compare the radial
density distribution $r^{2}\rho(r)$ of the LQ and HQ models to that
predicted by other theories for the Kr atom.
Both the TF1/5W and TFW models predict smooth
and structureless radial density profiles.
Using the full Coulombic potential, all the LR-based models 
can predict an incipient shell structure for heavy
atoms ($Z\gtrsim30$), and these results are typical.
Since the potential is certainly far beyond the LR regime,
these qualitative results with some suggestion
of shell structure are about as good as could be hoped for.
The surprisingly good total energies given in Table \ref{table:1}
for the TF1/5W model and the LR-based models shows that
averaged thermodynamic properties are less sensitive to errors
in the KP than is the density profile. The difference
in the results for the LQ, HQ and CAT models indicates that the
LR-based OF theory 
is being used outside its range of validity. As shown below, we gain
a significant improvement by using the AILPS to deal with these difficulties.

\subsection{\textit{Ab initio} local pseudopotential calculations}

As discussed earlier, the use of pseudopotentials
in nonlocal LR-based OF-DFT can improve the
accuracy of the theory because the weaker pseudopotential is
more nearly in the LR regime, where the theory is designed to be accurate.
Our proposed HQ and LQ models can be used with any existing AILPS. However,
since we want to assess the performance of these models for a wide class of atomic
systems, we describe here a new method for
determining reasonable AILPS for general atomic systems.
These pseudopotentials will be used in all our calculations and
can be transfered to other molecular and solid
state environments, but we expect (and find
in cases where comparison can be made) little change if other 
reasonable AILPS are used.

\begin{figure}[t]

\caption{\label{Fig.5.6} The smooth target density $\tilde{\rho}_{v}(r)$
from Eq.\ (\ref{eq:4.7.3}), with parameters given in Table \ref{table:4.1}
for the Si pseudoatom used in the inverse-KS process, and the valence
density $\rho_{v}(r)$ predicted by the LQ and HQ models using
the $V_{ps}(r)$ (see Fig. \ref{Fig.4.3}) corresponding to $\tilde{\rho}_{v}(r)$.
The arrow indicates the location of $r_{c}$.}

\includegraphics[%
  scale=0.7]{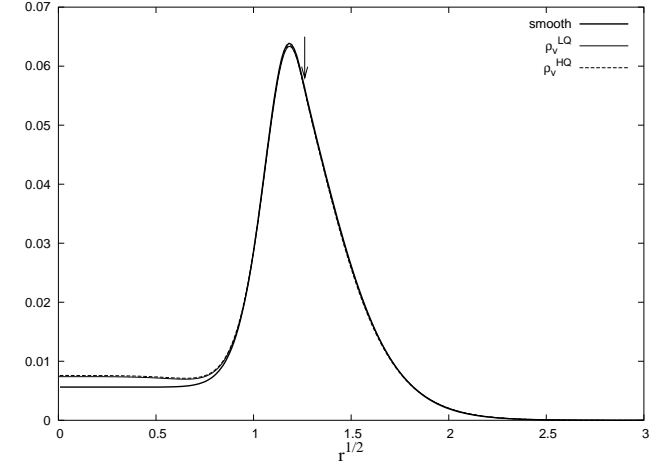}
\end{figure}

\begin{figure}[t]

\caption{\label{Fig.4.3} The AILPS $V_{ps}(r)$ for Si
generated by the target density $\tilde{\rho}_{v}(r)$ in Fig. \ref{Fig.5.6}.
}

\includegraphics[%
  scale=0.7]{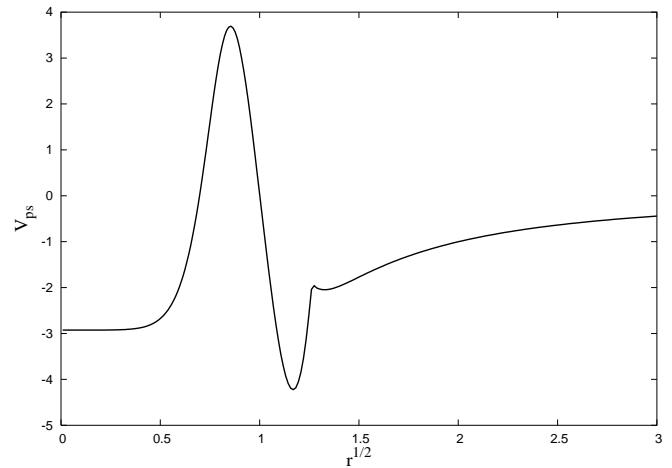}
\end{figure}

Because of the one-to-one mapping between the effective one-body potential
acting on a system of $N$ electrons and the electron density in the
ground-state configuration, it is possible to obtain a unique local
one-body potential that generates a given target density $\tilde{\rho}({\bf r})$ by using a KS
orbital-based method in an inverse way \cite{WangParr}. To construct an AILPS
for a given atom we separate the total electron density $\rho({\bf r})$ into a
``core density'' $\tilde{\rho}_{c}({\bf r})$, which is supposed not to
vary significantly in other molecular or solid state environments,
and the target ``valence density'' $\tilde{\rho}_{v}({\bf r})$ where
\begin{equation}
\rho({\bf r})=\tilde{\rho}_{v}({\bf r})+\tilde{\rho}_{c}({\bf r}).
\label{eq:4.7.1}
\end{equation}

Because DFT requires only the electron density, we can take
a more general view of what is meant by the core and valence components
than is used in most orbital-based methods. 
Here, we directly construct a smooth target valence density for the
$N_{v}=N-N_{c}$ valence electrons, with $N_{c}$
chosen to be the number of electrons in the noble gas configuration.

Our proposed target valence density $\tilde{\rho}_{v}({r})$ for atoms equals
the full KS density $\rho_{KS}(r)$ outside a core of radius $r_{c}$,
and is designed to be small and slowly varying inside $r_{c}$. The
functional form we take is
\begin{equation}
\tilde{\rho}_{v}(r)=\left\{ \begin{array}{ll}
t\rho_{KS}(r_{c})+a_{0}r^{q}\exp[-r^{p}(a_{_{1}}+a_{_{2}}r^{2})]\bigskip & \; r\leq r_{c}\\
\rho_{KS}(r) & \; r>r_{c}\end{array}\right.
\label{eq:4.7.3}
\end{equation}
Figure \ref{Fig.5.6} gives an example of $\tilde{\rho}_{v}({r})$
for Si that will be discussed in more detail below. We find most results are insensitive
to the details of our fitting procedure. Parameter
values for a variety of atomic systems are given in the Appendix in Table \ref{table:4.1} along
with the physical and technical considerations that guided our choice of
this particular form for $\tilde{\rho}_{v}({r})$. The Appendix also discusses
some of the general issues that arise in using these atomic AILPS in other environments.

\begin{figure}[t]

\caption{\label{Fig.5.7} (Color online) Radial valence density $r^{2}\rho_{v}(r)$ of the Si atom using
the KS method and various models using AILPS. Parameters used
for constructing this reference system are shown in Table \ref{table:4.1}.
The arrow indicates the location of $r_{c}$. Inset: The corresponding 
radial total density $r^{2}\rho(r)$, which is dominated by the core component
for $r<r_c$.}

\includegraphics[%
  scale=0.7]{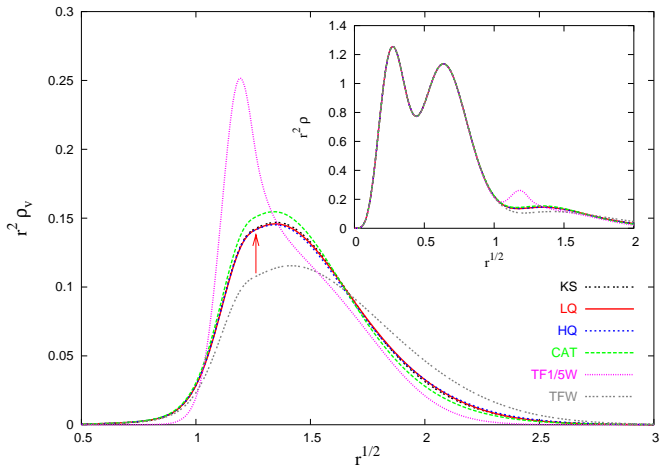}
\end{figure}

The local pseudopotential is directly related to
the effective one body potential that reproduces $\tilde{\rho}_{v}({\bf r})$ exactly
when using the full KS theory. Following previous work
\cite{Zhou,BWang}, for a given $\tilde{\rho}_{v}({\bf r})$, the inverse-KS
equations are solved to get the effective one-body screened potential
$V_{scr}({\bf r})$. The desired \textit{ab initio} local pseudopotential
$V_{ps}({\bf r})$ is then obtained by subtracting the Hartree
potential and the exchange-correlation potential:
 \begin{equation}
V_{ps}({\bf r})=V_{scr}({\bf r})-V_{H}({\bf r};[\tilde{\rho}_{v}])
-V_{xc}({\bf r};[\tilde{\rho}_{v}]).
\label{eq:4.7.2}\end{equation}
This relatively expensive procedure to determine $V_{ps}({\bf r})$
requires the use of orbitals. However it needs to be done only once
for each atom, and the resulting $V_{ps}({\bf r})$ can then be used in
a variety of other environments if the atomic core densities remain essentially constant.

Once suitable $V_{ps}({\bf r})$ have been determined (by this or other means)
they can be incorporated in $V_{ext}({\bf r})$ in different ways,
depending on the particular system of interest. OF-DFT theory
can then be used to determine the valence density
$\rho_{v}({\bf r})$ in direct analogy to the
all-electron calculations for the full atomic potential in Eq.\ (\ref{eq:51}):
\begin{equation}
\begin{array}[b]{ll}
\left\{ -\frac{1}{2}{\bf \bigtriangledown}^{2}+V_{H}({\bf r};[\rho_{v}])+V_{xc}({\bf r};[\rho_{v}])\medskip\right.\\
\;\;\;\;\;\;\;\;\;\;\;\;\;\;\;\left.+\medskip V_{ext}({\bf r})+V_{P}({\bf r};[\rho_{v}])\right\} \psi_{v}({\bf r})
=\mu\psi_{v}({\bf r})
\end{array}\label{eq:54}
\end{equation}
 where 
\begin{equation}
\rho_{v}({\bf r})=|\psi_{v}({\bf r})|^{2}.\label{eq:55}
\end{equation}

A simple and direct test of OF-DFT is to use Eqs.\ (\ref{eq:54}) and (\ref{eq:55})
for the same atomic system for which $V_{ps}({\bf r})$ was constructed.
Thus we take $V_{ext}({\bf r})=V_{ps}({\bf r})$
for a given atom as input data in Eq.\ (\ref{eq:54}).
The valence density $\rho_{v}({\bf r})$ predicted
by the OF KPs is determined from Eqs.\ (\ref{eq:54}) and (\ref{eq:55}),
and can be directly compared to the exact target density $\tilde{\rho}_{v}({\bf r})$
for this atomic system given by the full KS theory.

This is illustrated in Fig.\  \ref{Fig.5.6}, which shows the input 
target valence density $\tilde{\rho}_{v}(r)$ for Si used in the inverse-KS process.
The total KS density $\rho_{KS}(r)$ equals $\tilde{\rho}_{v}(r)$ for
$r \geq r_{c}$, indicated by the arrow in Fig.\  \ref{Fig.5.6}, and
 then increases rapidly for $r<r_{c}$, reaching a very large value at the nucleus,
$\rho_{KS}(0)=1754$. In contrast, the proposed target
valence density $\tilde{\rho}_{v}(r)$
remains small and relatively slowly varying inside the core, with $\tilde{\rho}_{v}(0)=0.005621$.

Also shown in Fig.\  \ref{Fig.5.6} are the predicted valence densities
$\rho_{v}(r)$ for Si given by the LQ and HQ models.
Because of the relatively weak $V_{ps}(r)$ and slowly varying valence density,
both the LQ and HQ KP models predict results very close to those given by
the exact KS treatment of the kinetic energy, and perform markedly better than
they did for the all-electron calculations using the full Coulomb potential.
Fig.\ \ref{Fig.4.3} shows the corresponding AILPS
generated by the inverse KS procedure. It is much smaller in the core region
than the full atomic potential and more likely to be accurately treated by LR-based
methods.
\begin{figure}[t]
\caption{\label{Fig.5.10} (Color online) Same as in Fig.\ \ref{Fig.5.7} but 
for the Ar atom.}
\includegraphics[%
  scale=0.7]{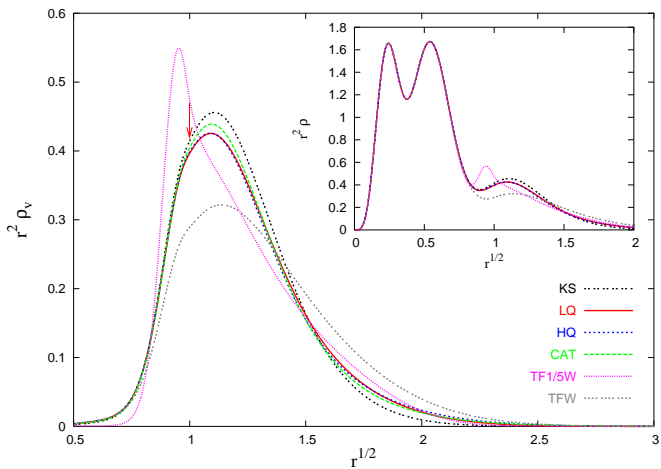}
\end{figure}

The radial valence density $r^2\rho_{v}(r)$ of the Si and Ar atoms
predicted by the various methods are shown in Figs.\ \ref{Fig.5.7}
and \ref{Fig.5.10}. The consistency of our OF theory when pseudopotentials are used is
illustrated by the similarity of the density predicted by the
HQ and LQ models. The slight deviations from the KS-DFT
results for Ar in the valence region in Fig.\ \ref{Fig.5.10} are among the largest
we encountered for all atoms tested, and could be due to the relatively
large number of valence electrons (8) compared to
core electrons (10). Further errors may arise from the
LFWV approximation in Eq.\ (\ref{eq:3.6.2}).

Once $\rho_{v}({\bf r})$ has been determined using OF-DFT, it can be added
to the known input core density $\tilde{\rho}_{c}({\bf r})$ to obtain the predicted total density,
since the basic assumption of our AILPS is that the core density remains unchanged in 
different chemical environments. The core density
is defined in Eqs.\ (\ref{eq:4.7.1}) and (\ref{eq:4.7.3}). The inset in Fig.\ \ref{Fig.5.7}
shows that the HQ and LQ treatment of the valence density for
the Si atom does not produce noticeable errors in the total density,
as expected, and the shell structure remains in excellent agreement with
the full KS-DFT calculations.
 
As a further test of OF-DFT we can compare the energy for the valence density
given by the various methods to the exact
valence energy for the target valence density determined by the inverse KS method.
Table \ref{table:5.3} gives the valence energy
values for the KS-DFT, and the LQ and the HQ models. As can been seen,
both the LQ and HQ models give very good agreement with KS-DFT and 
perform significantly better than the other models. These results show that for this class
of relatively weak pseudopotentials the OF treatment of the KP is quite satisfactory.
\begin{table}[t]

\caption{\label{table:5.3} The total valence energy $E_{v}[\rho_{v}]$ 
using the KS method, the LQ and HQ models, the CAT model, and the
TF$\lambda$W models. 
MAE, the mean absolute error (relative to the KS method) of various
OF models are given at the bottom of their respective columns.  
Parameters used in
Eq.\ (\ref{eq:4.7.3}) for such systems are given in Table \ref{table:4.1}.}

\begin{tabular}{ccccccc}
\hline 
&
 KS&
 LQ&
 HQ&
CAT&
TF1/5W&
TFW\tabularnewline
\hline
Be&
 $-0.9914$&
 $-0.8955$&
 $-0.8950$&
 $-0.9583$&
 $-1.214$&
 $-0.7786$\tabularnewline
\hline
C&
 $-6.134$&
 $-6.080$&
 $-6.100$&
 $-6.345$&
 $-7.761$&
 $-5.266$\tabularnewline
\hline
N&
 $-11.04$&
 $-11.06$&
 $-11.09$&
 $-11.32$&
 $-13.93$&
 $-9.462$\tabularnewline
\hline
O&
 $-18.01$&
 $-18.09$&
 $-18.09$&
 $-18.19$&
 $-22.44$&
 $-15.30$\tabularnewline
\hline
Si&
 $-3.771$&
 $-3.738$&
 $-3.750$&
 $-3.869$&
 $-4.467$&
 $-3.350$\tabularnewline
\hline
P&
 $-6.474$&
 $-6.432$&
 $-6.455$&
 $-6.582$&
 $-7.385$&
 $-5.756$\tabularnewline
\hline
S&
 $-10.20$&
 $-10.10$&
 $-10.14$&
 $-10.24$&
 $-11.27$&
 $-9.023$\tabularnewline
\hline
Ar&
 $-21.37$&
 $-20.84$&
 $-20.91$&
 $-20.85$&
 $-22.40$&
 $-18.56$\tabularnewline
\hline
MAE&
 &
 $0.119$&
 $0.103$&
 $0.184$&
 $1.610$&
 $1.312$ \tabularnewline
\hline
\end{tabular}
\end{table}

\begin{figure}[t]

\caption{\label{Fig.5.11} (Color online) Ionization energies (shown in Hartree) of the first and 
the second row atoms using the full KS method, and various models using AILPS.
The mean absolute errors (relative to the full KS method) of various
models using AILPS are: KS (0.1 eV), LQ (1.8 eV), HQ (2.1 eV), CAT (3.9 eV),
TFW (3.1 eV), and TF1/5W (4.3 eV). Ionization energies using the TF1/5W model
are not shown in the figure due to its relatively poor performance.} 
\includegraphics[%
  scale=0.7]{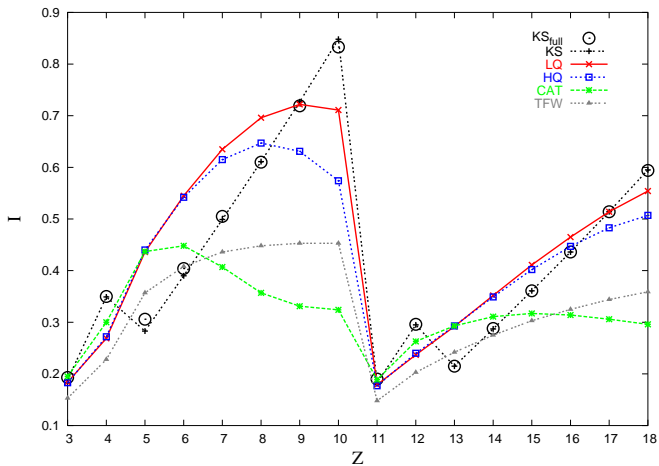}
\end{figure}

To test of the transferability of the present AILPS, we
also performed calculations for positive ions. Fig.\ \ref{Fig.5.11} shows
that the ionization
energies of various  atoms calculated using KS-DFT and the full
atomic densities and those from the KS-DFT 
using the valence densities with AILPS are very similar. 
Therefore, the present AILPS are quite transferable to these 
positive ions.  

The ionization energy for models using the AILPS is obtained
by subtracting the valence energies for systems with $N_v$ and $N_v-1$ electrons.
Since the core electrons are assumed to be unaltered in different chemical 
environments, the atomic core energy is a constant that 
cancels here or in other similar applications to molecules and solids. 
Limitations of the LQ and HQ models are more evident here,
but they do capture the overall periodicity of the ionization 
energies well, and perform significantly better than the other models.

\section{Conclusion}

In summary, we propose two nonlocal OF KPs that satisfy exact limits
for small and large wavevector perturbations and reproduce the exact
LR function in the homogeneous limit. These are the same limits that
several current KEDFs are designed to satisfy. However, because of
the more local nature of the KP, it is much
easier to satisfy these conditions for the KP than for the KEDF,
and there may be other physical and technical advantages arising
from use the more local KP.

In general, there is no reason
to believe that any LR-based OF-DFT should work well for arbitrary
systems where the model potentials are far beyond the LR
regime. However, most chemical processes involve
changes of valence electron densities, which can
often be described by a weak AILPS. Thus the use
of a LR-based OF-KP together with AILPS for atomic systems at least
seems well justified. The small and relatively slowly varying $\rho_{v}(r)$
also provides some justification for our use of the local
FWV $k_{F}({\bf r})$ in Eq.\ (\ref{eq:3.6.2}).

When the AILPS is used, the valence densities given by the LQ and HQ KPs are
close to those given by the KS method. Thus
the particular integration pathway used to get the total energy value
becomes unimportant. The simple pathway in Eq.\ (\ref{eq:6})
is especially useful, since no coupling parameter integration is needed.

The proposed models are not only conceptually simple, but also exact
for a model system with a weak potential and a slowly-varying density.
The appearance of the atomic shell structure was found to be very sensitive
to the accuracy of the proposed KPs. The LR-based LQ and
HQ KPs give at best only qualitative indications of shell structure for full atomic systems,
though total energies are surprisingly good.
Still better results for atoms and ions can be found by focusing on the valence density
as determined by a relatively weak AILPS.
While these results seem promising,  improved KPs are needed
and further investigation
is required to see if these ideas can be usefully applied to
other relevant systems like molecules and solids.
Some initial results along these lines will be reported elsewhere.

\begin{acknowledgments}
This work has been supported by the NSF Grant CHE01-11104, and by
the NSF-MRSEC at the University of Maryland under Grant No. DMR 00-80008.
J.D.C. acknowledges the support from the UMCP Graduate School Fellowship,
the IPST Alexander Fellowship, and the CHPH Bolck Grant Supplemental
Fellowship. We are grateful to Prof. Emily Carter and members of her group 
for many helpful discussions and comments on an earlier version of this paper.
\end{acknowledgments}

\appendix*

\section{Ab Initio Local Pseudopotentials}

\begin{table}[t]

\caption{\label{table:4.1} Parameters used in Eq.\ (\ref{eq:4.7.3a}) for the
target valence density of various atoms. Here, $p=q=6$, and $t=0.1$ are
used for all systems. The $V_{ps}(r)$ generated from these parameterized
$\tilde{\rho}_{v}(r)$ can then be used in OF-DFT.}

\begin{tabular}{cccccc}
\hline 
&
$N_{v}$&
 $a_{0}$&
 $a_{1}$&
 $a_{2}$&
 $r_{c}$\tabularnewline
\hline
Li&
$1$&
$2.983\times10^{-4}$&
$0.05260$&
$-6.560\times10^{-3}$&
$2.135$\tabularnewline
\hline
Be&
$2$&
 $0.02655$&
 $0.7078$&
 $-0.2130$&
 $1.370$\tabularnewline
\hline
B&
$3$&
$0.8032$&
$5.855$&
$-3.531$&
$0.9714$\tabularnewline
\hline
C&
$4$&
 $11.53$&
 $30.17$&
 $-31.20$&
 $0.7429$\tabularnewline
\hline
N&
$5$&
 $98.80$&
 $112.8$&
 $-180.2$&
 $0.5978$\tabularnewline
\hline
O&
$6$&
 $592.7$&
 $338.9$&
 $-779.8$&
 $0.4981$\tabularnewline
\hline
F&
$7$&
$2750$&
$871.4$&
$-2744$&
$0.4256$\tabularnewline
\hline
Ne&
$8$&
$1.052\times10^{4}$&
$1993$&
$-8267$&
$0.3707$\tabularnewline
\hline
Na&
$1$&
$5.234\times10^{-5}$&
$9.840\times10^{-3}$&
$-6.610\times10^{-4}$&
$2.904$\tabularnewline
\hline
Mg&
$2$&
$9.805\times10^{-4}$&
$0.04445$&
$-5.002\times10^{-3}$&
$2.233$\tabularnewline
\hline
Al&
$3$&
$6.164\times10^{-3}$&
$0.1273$&
$-0.02056$&
$1.861$\tabularnewline
\hline
Si&
$4$&
 $0.02942$&
 $0.3119$&
 $-0.06861$&
 $1.593$\tabularnewline
\hline
P&
$5$&
 $0.1149$&
 $0.6864$&
 $-0.1979$&
 $1.390$\tabularnewline
\hline
S&
$6$&
 $0.3841$&
 $1.390$&
 $-0.5100$&
 $1.231$\tabularnewline
\hline
Cl&
$7$&
$1.133$&
$2.632$&
$-1.201$&
$1.103$\tabularnewline
\hline
Ar&
$8$&
 $3.018$&
 $4.714$&
 $-2.624$&
 $0.9985$\tabularnewline
\hline
\end{tabular}
\end{table}

As discussed above, our proposed target valence density
for atoms has the following form:
\begin{equation}
\tilde{\rho}_{v}(r)=\left\{ \begin{array}{ll}
t\rho_{KS}(r_{c})+a_{0}r^{q}\exp[-r^{p}(a_{_{1}}+a_{_{2}}r^{2})]\bigskip & \; r\leq r_{c}\\
\rho_{KS}(r) & \; r>r_{c}\end{array}\right.
\label{eq:4.7.3a}
\end{equation}
Here $p$ and $q$ are taken as even integers. The
larger they are, the smaller and more slowly varying is the valence density near
$r=0$ but the sharper is the peak near the core radius $r_c$.
As a compromise, we take here $p=q=6$, which generates relatively slowly varying
local pseudopotentials $V_{ps}(r)$.
For applications in different environments,
such as molecules or crystals, the core size $r_{c}$ has to be small
to maintain transferability of the atomic core density. For this reason,
we force $\tilde{\rho}_{v}(r=0)$ to be small by taking a small $t$.
If $t=0$, the strict vanishing of the valence density near
the nucleus would require a very repulsive $V_{ps}$, which is certainly
undesirable for the LR-based OF-DFT. However, if $t$
is too large, there will exist a long oscillatory tail outside the
core in the corresponding $V_{ps}(r)$. This is an undesirable feature
for transferability to other environments,
as will be discussed below. These two points constrain the value
of $t$ and we use here $t=0.1$ for all the atomic systems
considered.

The four parameters $a_{0}$, $a_{1}$, $a_{2}$, and $r_{c}$ are
determined by requiring continuity of the function $\tilde{\rho}_{v}(r)$
and its first two derivatives at $r=r_{c}$, and by satisfying the
normalization condition: 
\begin{equation}
N_{v}=4\pi\int\tilde{\rho}_{v}(r)r^{2}dr.\label{eq:4.7.4}
\end{equation}
Here we used
the standard noble gas cores to determine $N_v=N-N_c$, though other
choices could in principle be made.

To construct our local pseudopotential $V_{ps}(r)$ for atoms, we
first solve the KS equations for an atom with the full Coulomb potential.
With the KS density $\rho_{KS}(r)$ and $N_c$ determined, this construction
ensures that as $r\rightarrow r_{c}$, 
the associated core density smoothly approaches zero as
$O(\mid r-r_{c}\mid^{3})$.

\begin{figure}[t]

\caption{\label{Fig.4.4} The inverse KS procedure generates
very small oscillations in the tail of $rV_{ps}(r)$ for Si (shown here) and other atoms.
The two points where $rV_{ps}(r)=-4$ for Si are
$r_{1}=2.336$ and $r_{2}=4.576$ or $r_{2}^{1/2}=2.139$. The arrow indicates the location of $r_{2}$.
To achieve good transferability (see text),
this $rV_{ps}(r)$ is modified by setting $rV_{ps}(r)=-4$ 
when $r \geq r_{2}=4.576$. See Fig.\ \ref{Fig.4.3} for a large scale view.}

\includegraphics[%
  scale=0.7]{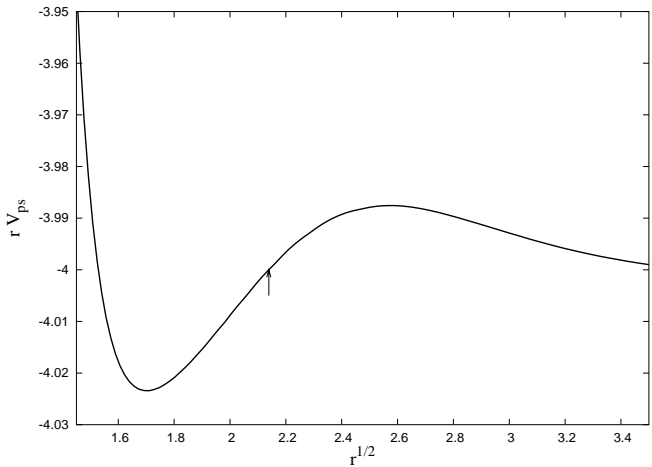}
\end{figure}

After generating the parameterized target valence density $\tilde{\rho}_{v}(r)$,
the set of inverse-KS equations are solved to obtain the corresponding
one-body screened potential $V_{scr}(r).$ In principle, the AILPS $V_{ps}(r)$
is then given by Eq.\ (\ref{eq:4.7.2}).
Using this, we find with an acceptable $N_{c}$ that
essentially the same density profiles outside $r_c$
are predicted by the LQ and HQ models for a wide range of 
choices of $p$, $q$, and $t$. This would be expected if most features of the resulting
set of model potentials are within the different regimes accurately described
by the OF KP. These results clearly show that OF-DFT can give
accurate results for this class of physically relevant  and relatively weak and slowly varying
model potentials. 

However, we found that the $rV_{ps}(r)$ constructed in this way for group III
to group VIII elements deviates from $-Z_v$ outside the core by a small
and long range oscillation. For the $2s^{2}2p^{x}$
atoms, the maximum amplitude of the oscillation is about $0.18$. For the
$3s^{2}3p^{x}$ atoms, it ranges from $0.003$ for Al to $0.037$
for Ar. Similar oscillatory tails were observed by Wang and Stott
\cite{BWang}, and are thought to arise from the inability of
a local pseudopotential to represent both $s$ and $p$
orbitals of the corresponding nonlocal pseudopotentials. These oscillations
are so small in magnitude that they
have almost no effect on the density profile or energy of atoms, but they
can cause transferability problems when used for solids \cite{Bylander}.
Following Wang and Stott's approach,
we replace $V_{ps}(r)$ by $-Z_v/r$ at the largest point
where $rV_{ps}(r)=-Z_v$ (see Fig.\ (\ref{Fig.4.4})). 
The new $V_{ps}(r)$ can then be used for both atomic and solid state calculations.

In other chemical environments, such molecules
and solids, the predetermined AILPS $V_{ps}({\bf r})$ centered at
each nucleus are regarded as input data, and the valence densities
for other systems can be calculated by OF-DFT. Results can be checked
by using the full KS-DFT. For example, for molecules and solids, the
external potential $V_{ext}({\bf r})$ is a linear combination of
the local atomic pseudopotentials centered at each ion position ${\bf R}_{I}$:
\begin{equation}
V_{ext}({\bf r})=\sum_{I}V_{ps}({\bf r}-{\bf R}_{I})\label{eq:4.7.5}
\end{equation}

The valence density can then be determined from Eqs.\ (\ref{eq:54})
and (\ref{eq:55}), and the total energy of valence electrons is
\begin{equation}
E_{v}[\rho_{v}]=T_{s}[\rho_{v}]+E_{H}[\rho_{v}]+E_{xc}[\rho_{v}]+\int\rho_{v}({\bf r})V_{ext}({\bf r})d{\bf r}
\label{eq:4.7.8}
\end{equation}
For the proposed LQ  and HQ KPs, the value of $T_{s}[\rho_{v}]$ in Eq.\ (\ref{eq:4.7.8})
is determined by Eqs.\ (\ref{eq:6}) or (\ref{eq:2.3.4}) with $\rho({\bf r})$ 
replaced by $\rho_{v}({\bf r})$.

\bibliographystyle{jcp}
\bibliography{basename of}

\begin{references}
\bibitem{CATcalculations} Instead of solving the second-order differential equation
numerically for the corresponding weight function $\omega(x)$ (see Eq. (14) of Ref.\ \cite{GAC}),
we used the parametrized form in the appendix of Ref.\ \cite{GAC}. Although
the parametrization was made for $\beta=-1/2$
(note the sign change of $\beta$ in Ref.\ \cite{GAC}), the $\beta$
dependence of $\omega(x)$ is small for small $\beta$, as argued
in Ref.\ \cite{GAC2}. Our results for the CAT model using this parametrized $\omega(x)$
are very close to their reported results.
\end{references}

\end{document}